\documentstyle[aps,prb,multicol,epsf,rotate]{revtex}

\begin{document}
\draft

\title{\rule[0.375in]{0in}{0.375in}
Local Time-Reversal Symmetry Breaking in $d_{x^2-y^2}$ Superconductors}

\author{M.~J.~Graf$^{\, 1}$, A.~V.~Balatsky$^{\, 1,2}$, and J.~A.~Sauls$^3$}
\address{$^1$Center for Materials Science, $^2$Theory Division\\
Los Alamos National Laboratory, Los Alamos, New Mexico 87545\\
$^3${Department of Physics \& Astronomy, Northwestern University, 
Evanston, Illinois 60208}}

\date{\today}
\maketitle

\begin{abstract}
We show that an isolated impurity in a spin singlet $d_{x^2-y^2}$ 
superconductor generates a {$d_{xy}$}
order parameter with locally broken time-reversal symmetry.
The origin of this effect is a coupling between the $d_{x^2-y^2}$
and the $d_{xy}$ order parameter induced by spin-orbit scattering off
the impurity. The signature of locally broken 
time-reversal symmetry is an induced orbital 
charge current near the impurity, which
generates a localized magnetic field in the vicinity of the impurity.
We present a microscopic theory for the impurity induced  $d_{xy}$ component,
discuss its spatial structure as well as the pattern of induced 
current and local magnetic field near the localized impurity spin.
\end{abstract}

\pacs{{PACS numbers: 74.25Bt, 74.62Dh} \hfill{LA-UR: 99-1349}}

\begin{multicols}{2}

There is now strong evidence to support the identification 
of the superconducting state of many of the high T$_c$ cuprates 
with a spin-singlet pairing amplitude having ``d-wave'' orbital
symmetry, or more precisely $d_{x^2-y^2}$ symmetry.\cite{van95}
This phase preserves time-reversal (${\cal T}$) symmetry, 
{but changes sign under reflection along the [110] 
and [$\bar{1}$10] mirror planes, as well as $\pi/2$-rotations
in a tetragonal crystal.}
As a consequence $d_{x^2-y^2}$ pairing correlations 
are particularly sensitive to scattering of 
quasiparticles on the Fermi surface.
In this article we show that an isolated impurity 
in a spin singlet $d_{x^2-y^2}$ superconductor generates a 
complex $d_{xy}$ order parameter {(OP)}
with locally broken ${\cal T}$ symmetry; the
signature of this effect is an induced orbital charge 
current near the impurity and a localized magnetic field 
in the vicinity of the impurity.

Atomic scale impurities, or defects,
scatter conduction electrons which leads to local suppression
of the superconducting {OP} (pair-breaking) near the impurity.
The mechanism responsible for pair-breaking is the formation of
quasiparticle states near the Fermi level which are
bound to the impurity by Andreev scattering.\cite{thuneberg81b}
The corresponding
reduction in the spectral weight of the pair condensate is 
responsible for pair-breaking.
The existence of quasiparticle states near
the Fermi level can also lead to 
local Fermi-surface instabilities and mixing of
order parameters with different symmetry.\cite{buc95b}
Low temperature
phase transitions associated with a secondary {OP}
may provide new information on the mechanism(s) for pairing in 
unconventional superconductors,\cite{fog97a} while
impurity-induced {\sl mixing} of the $d_{x^2-y^2}$ {OP}
with an {OP} of different symmetry can provide direct
information on the atomic structure of the impurity.\cite{balatsky98}

Recent transport experiments report evidence for
low temperature phases associated with a secondary {OP}
at surfaces ($T_s\simeq 8\,\mbox{K}$).\cite{cov97}
This was interpreted
in terms of a surface phase transition to a $d_{x^2-y^2}+is$ state with
spontaneously broken ${\cal T}$-symmetry.\cite{fog97a}
Bulk transport measurements 
on $\rm Bi_2 Sr_2 Ca Cu_2 O_{8+\delta}$ (Bi-2212)
show a pronounced drop in the thermal conductivity at
$T^* \approx 150\, {\rm mK} \ll T_c \approx 80\, {\rm K}$ 
in Ni doped Bi-2212.\cite{movshovich98}
This anomaly was interpreted as
the signature of a second superconducting phase with a fully gapped 
quasiparticle spectrum and a mixed symmetry {OP}
of the form $d_{x^2-y^2} +id_{xy}$.
This phase {was} proposed to arise from a coupling of
the orbital momentum of the conduction electron with the spin of
the magnetic impurity,
${\cal H}_{\mbox{\tiny so}}=
\int d{\bf r}\,\psi^{\dag}_{\alpha}({\bf r})\,
v(r)\,{\bf L}_{\mbox{\tiny orbit}}\cdot
{\bf S}_{\mbox{\tiny imp}}
\psi_{\alpha}({\bf r})$.\cite{balatsky98}
Measurements of the spin-orbit coupling energy for Ni$^{2+}$ ions
indicate that it is a few percent of the nonmagnetic and
exchange channels.\cite{yosida96}
In this model the $id_{xy}$ {OP} is induced at a temperature
above the second phase transition, $T^* <  T \leq T_c$; the
low temperature transition is argued to be ordering of 
the impurity-induced ``patches'' of the local $d_{x^2-y^2} \pm id_{xy}$
order.\cite{balatsky98} Above $T^*$ patches with randomly fluctuating 
internal phase destroy the long range order, $\langle d_{xy}\rangle = 0$,
but preserve $\langle |d_{xy}|^2\rangle \neq 0$. Thus,
the local structure associated with $d_{x^2-y^2} \pm id_{xy}$ symmetry
near a magnetic impurity should be observable at temperatures 
{well above $T^*$.}
{The electronic and magnetic structure near an impurity located near the
surface of a superconductor can now be studied with atomic resolution at 
low temperature by scanning tunneling microscopy (STM),
opening a new window for local probes.\cite{yaz97,hudson99}
}

In this article we investigate theoretically the local structure of
the {OP} and the current distribution in 
the neighborhood of an atomic impurity
within the $d_{x^2-y^2}$ model for the high T$_c$ cuprates.
We present new analytical and numerical results for the
local $d_{x^2-y^2} \pm i d_{xy}$ {OP} that develops near a magnetic impurity,
{which has attracted new theoretical interests.\cite{grimaldi99,simon99}}
Our approach follows closely the theory developed in the
late 70's for ions in superfluid $^3$He,\cite{rainer77,thuneberg81b}
and later adapted to study the properties of impurities
in heavy fermion superconductors.\cite{choi89,choi90}
The theory of impurity scattering in superconductors
can be formulated to quasiclassical accuracy as
an expansion in $\sigma/\xi_0$, where $\sigma$ is the (linear
in 2D) cross-section of the impurity for scattering of normal-state 
quasiparticles at the Fermi surface, and $\xi_0=v_f/\pi \Delta_0$ is the
superconducting coherence length. This ratio is typically
very small in low T$_c$ superconductors and in superfluid $^3$He, but 
{may be as big as $0.2$ for strong scatterers in high 
T$_c$ superconductors.}

We start from Eilenberger's transport equation,
for the matrix propagator in particle-hole/spin space,
{$\hat{g}({\bf p}_f,{\bf r};\epsilon_n)$}.
The diagonal element of the propagator determines the local density of states
and local equilibrium current distribution, and the off-diagonal
elements are the components of the local pair amplitude.
Quasiparticle scattering off an isolated impurity 
is included through a {\sl source} term
{ on the {\it r.h.s} of the Eilenberger equation}.\cite{thuneberg81b}
The transport equation can be linearized to leading
order in $\sigma/\xi_0$ for distances $r\gg\sigma$ from the 
impurity. In this limit the Fourier 
transform of the linearized
transport equation reduces to,\cite{choi90}
\begin{eqnarray}\label{transport_eq}
\left[
 i\epsilon_n \hat\tau_3\!-\!\hat\Delta_b\!-\!\hat\sigma_{\text{imp}} ,
 \delta\hat g
\right]\!+\!{\bf q}\!\cdot\!{\bf v}_f \delta\hat g\!=\!\left[
 \hat t\!+\!\delta\hat\Delta\!+\!\delta\hat\sigma_{\text{imp}},
 \hat g_b
\right]\!.\!\!
\end{eqnarray}
The bulk propagator, $\hat g_b({\bf p}_f;\epsilon_n)
=-\pi [{ i \tilde\epsilon_n \hat\tau_3\!-\!\hat\Delta_b({\bf p}_f) }]/E$,
{order parameter},
$\hat\Delta_b({\bf p}_f)=\Delta_b({\bf p}_f)\hat{\tau}_1\,i\sigma_2$,
impurity scattering self-energy,
$\hat\sigma_{\text{imp}}$, and in-plane Fermi velocity,
${\bf v}_f$, are inputs to the linear response equations.
The energy denominator is given by
$E = (|\Delta_b({\bf p}_f)|^2 + \tilde\epsilon_n^2)^{1/2}$,
where $\tilde\epsilon_n=\epsilon_n\!+\!{i\over 4}{\rm Tr}\,\hat\tau_3
\hat\sigma_{\text{imp}}(\epsilon_n)$ is the renormalized
Matsubara frequency. The $\hat{t}$ matrix for the isolated impurity,
as well as the induced {OP}, 
$\delta\hat\Delta=[\delta\Delta_1\hat\tau_1+
\delta\Delta_2\hat\tau_2]i\sigma_2$, and self-energy,
$\delta\hat{\sigma}_{\text{imp}}$, enter the {\it r.h.s.}
of Eq.~(\ref{transport_eq}) as source terms.
{Here $\hat\tau_i$ and $\sigma_i$ are Pauli matrices in
particle-hole and spin space, respectively.}
The $\hat{t}$ matrix for the isolated impurity is given by
\begin{eqnarray}\label{tmatrix}
\hat t({\bf p}_f,{\bf p}_f';\epsilon_n) & = &
 \hat v({\bf p}_f,{\bf p}_f') + N_f \int d^2{\bf p}_f'' \,
 \hat v({\bf p}_f,{\bf p}_f'') 
\nonumber\\ && \times
 \hat g_b({\bf p}_f'';\epsilon_n) 
 \hat t({\bf p}_f'',{\bf p}_f';\epsilon_n)  \,,
\end{eqnarray}
{where $N_f$ is the 2D density of states at the Fermi energy per spin,
and}
$\hat v({\bf p}_f,{\bf p}_f')$ is the impurity
potential, which is evaluated in the forward scattering limit in 
Eq.~(\ref{transport_eq}); ${\bf p}_f'={\bf p}_f$.
We separate $\hat{v}$ into channels for nonmagnetic ($u$),
spin-spin exchange (${\bf m}=J{\bf S}_{\text{imp}}$), 
and spin-orbit scattering (${\bf u}_{\rm so}$) between the orbital 
momentum of the quasiparticle and the impurity spin,
${\bf S}_{\text{imp}}$, as well as the self-coupling {(${\bf w}_{\rm so}$)}
to the quasiparticle spin, $\hat{\bf S}$, \cite{barash97,spin_operator}
\begin{eqnarray}
\hat v({\bf p}_f,{\bf p}_f') &=& u({\bf p}_f,{\bf p}_f') + 
{\bf m}({\bf p}_f,{\bf p}_f') \cdot \hat{\bf S} +
\nonumber\\ &&
 i [ {\bf u}_{\rm so}({\bf p}_f,{\bf p}_f') \cdot {\bf S}_{\text{imp}} 
   +{{\bf w}_{\rm so}}({\bf p}_f,{\bf p}_f') \cdot \hat{\bf S} ]\hat\tau_3 .
\end{eqnarray}
The induced {OP} is self-consistently 
determined from the gap equation,
\begin{equation}
\delta\hat\Delta({\bf p}_f,{\bf q}) =
N_f\,T \sum_{\epsilon_n} \int\!\!d^2{\bf p}_f' V({\bf p}_f,{\bf p}_f') 
\hat f({\bf p}_f',{\bf q};\epsilon_n)\,,
\end{equation}
where $V({\bf p}_f,{\bf p}_f')$ is the pairing interaction and
$\hat{f}=[\delta f_1\hat\tau_1
+\delta f_2\hat\tau_2]i\sigma_2$ is the 
induced off-diagonal pair amplitude. We resolve
the pairing interaction into the dominant attractive
$d_{x^2-y^2}$ channel, and a secondary pairing channel with
$d_{xy}$ symmetry,
$V({\bf p}_f,{\bf p}_f') = V_{1} \eta_1(\phi) \eta_1(\phi')
 + V_{2} \eta_2(\phi) \eta_2(\phi')$,
where the eigenfunctions are $\eta_1(\phi) = \cos 2\phi$ and
$\eta_2(\phi) = \sin 2\phi$ for the two channels, respectively.
The dominant, attractive interaction is $V_{1}\equiv V_{x^2-y^2}$, 
and the subdominant interaction, $V_{2}\equiv V_{xy}$, 
may be either attractive or repulsive.
We neglect the $s$-wave pairing channel in order to simplify
the analysis, and we restrict our discussion
to the regime in which
the subdominant interaction, $V_{xy}$, is repulsive or
too weak to nucleate a bulk $d_{xy}$ {OP}.

Numerical calculations of the {OP} and current
distribution were carried out for an isolated impurity at
${\bf r} = 0$, with the $\hat t$ matrix source term of the form
$\hat t({\bf p}_f,{\bf p}_f;\epsilon_n) \delta({\bf r})$
in real space.
We modeled the position of the impurity to 
quasiclassical accuracy by replacing the delta function by
a smooth function, $\delta_{r_0}({\bf r})=
\frac{1}{\pi r_0^2} \exp(-r^2/r_0^2)$, of 
atomic width $r_0 = 0.1 \xi_0$. This model guarantees a
smooth cutoff in ${\bf q}$-space and faster convergence of the
Fourier integrals. For the computation reported here 
we also chose a subdominant pairing interaction 
corresponding to a bare subdominant transition 
temperature of $T_{c2}/T_{c1}=0.1$, which is well below
the threshold for bulk nucleation of a $d_{xy}$ order 
parameter.

The physical solution to the linearized transport 
equation (\ref{transport_eq}) is,\cite{choi90}
\begin{equation}\label{solution}
\delta\hat g({\bf p}_f,{\bf q};\epsilon_n) = 
\frac{ E \, \hat g_b + \pi Q }{2\pi( E^2 + Q^2 )}
 \big[\hat t + \delta\hat\Delta + \delta\hat\sigma_{\text{imp}} \, ,\, 
\hat g_b \big] ,
\end{equation}
where $Q={1\over 2}{\bf q}\cdot{\bf v}_f$. 
The induced charge current is also determined by
the $\hat t$ matrix and induced OP,
\begin{equation}\label{current_density}
\delta{\bf j}({\bf q})=
N_f T \sum_{\epsilon_n} \int\!\!d^2{\bf p}_f
{\bf v}_f\frac{2i\pi eQ\Delta_1}{E (E^2+Q^2)}
\left( t_2 + \delta\Delta_2 \right),
\end{equation}
where $t_2$ is the $\hat{\tau}_2$ component of $\hat{t}$.

We evaluate the $\hat t$ matrix
in {second-order Born approximation}, which is
adequate for weak scattering. More importantly, the Born
approximation generates the relevant coupling between the 
$d_{x^2-y^2}$ and $d_{xy}$ order parameters.
We also assume that the impurity potential
is short-ranged, so we retain only the scattering
amplitudes for the $s$-wave and $p$-wave channels, {\it i.e.}, 
$u({\bf p}_f,{\bf p}_f') \approx u_0 + u_1 {\bf p}_f \cdot {\bf p}_f'$, and
${\bf m}({\bf p}_f,{\bf p}_f') \approx (J_0 + J_1 {\bf p}_f \cdot {\bf p}_f'){\bf S}_{\text{imp}}$.
For the spin-orbit terms we approximate,
${\bf u}_{\rm so}({\bf p}_f,{\bf p}_f') \approx (\lambda_0 + \lambda_1 {\bf p}_f \cdot {\bf p}_f')
 {\bf p}_f \times {\bf p}_f'$, and
${\bf w}_{\rm so}({\bf p}_f,{\bf p}_f') \approx (w_0 + w_1 {\bf p}_f \cdot {\bf p}_f')
 {\bf p}_f \times {\bf p}_f'$.
The $\hat t$ matrix then has the general form
$\hat t = 
\big[ t_1 \eta_1 \hat\tau_1 + t_2 \eta_2 \hat\tau_2 \big] i \sigma_2
 + \big[ t_3 + {\bf t}\cdot \hat{\bf S} \big] \hat\tau_3$.
The important interference term is given by
\begin{equation}
t_2(\epsilon_n) = \pi  S_z N_f \tilde\lambda^2
\int \frac{d\phi}{2\pi} \frac{ \Delta_1 \eta_1^2(\phi) }{ 
 \sqrt{ |\Delta_1 \eta_1(\phi)|^2 + \tilde\epsilon_n^2 } } \,,
\end{equation}
with the {spin-orbit} parameter
$\tilde\lambda^2=u_0\lambda_1+u_1\lambda_0+J_0w_1+J_1w_0$,
{and the impurity spin $S_z$}.
The $t_2$ term generates 
a correction to the 
off-diagonal propagator with
$d_{x^2-y^2}$ (B$_{\text{1g}}$) symmetry 
and induces the $i d_{xy}$ (B$_{2g}$) OP 
near the impurity.

\begin{figure}
\noindent
\begin{minipage}{0.48\textwidth}
\epsfxsize=\hsize
\centerline{ \rotate[r]{\epsfbox{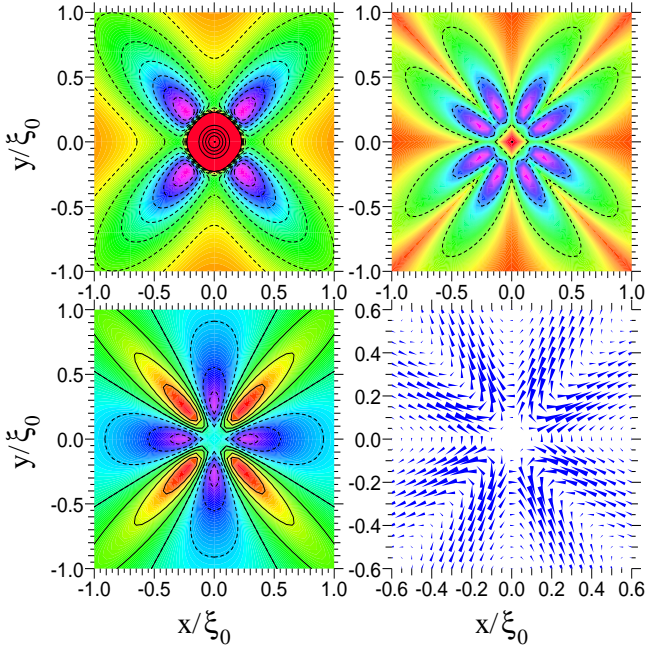}} }
\caption[]{Contour plots of the induced $i d_{xy}$ OP (top left),
the current density (top right), and the magnetic field (bottom left),
at $T=0.1 T_c$ for $T_{c2}/T_{c1}=0.1$ with no bulk scatterers.
Solid (dashed) lines are negative (positive) contour lines.
Bottom right: Field plot of the current density.
}\label{dXY}
\vspace{10pt}
\end{minipage}
\end{figure}

\begin{figure}
\noindent
\begin{minipage}{0.24\textwidth}
\epsfysize=0.99\hsize
\centerline{ \rotate[r]{\epsfbox{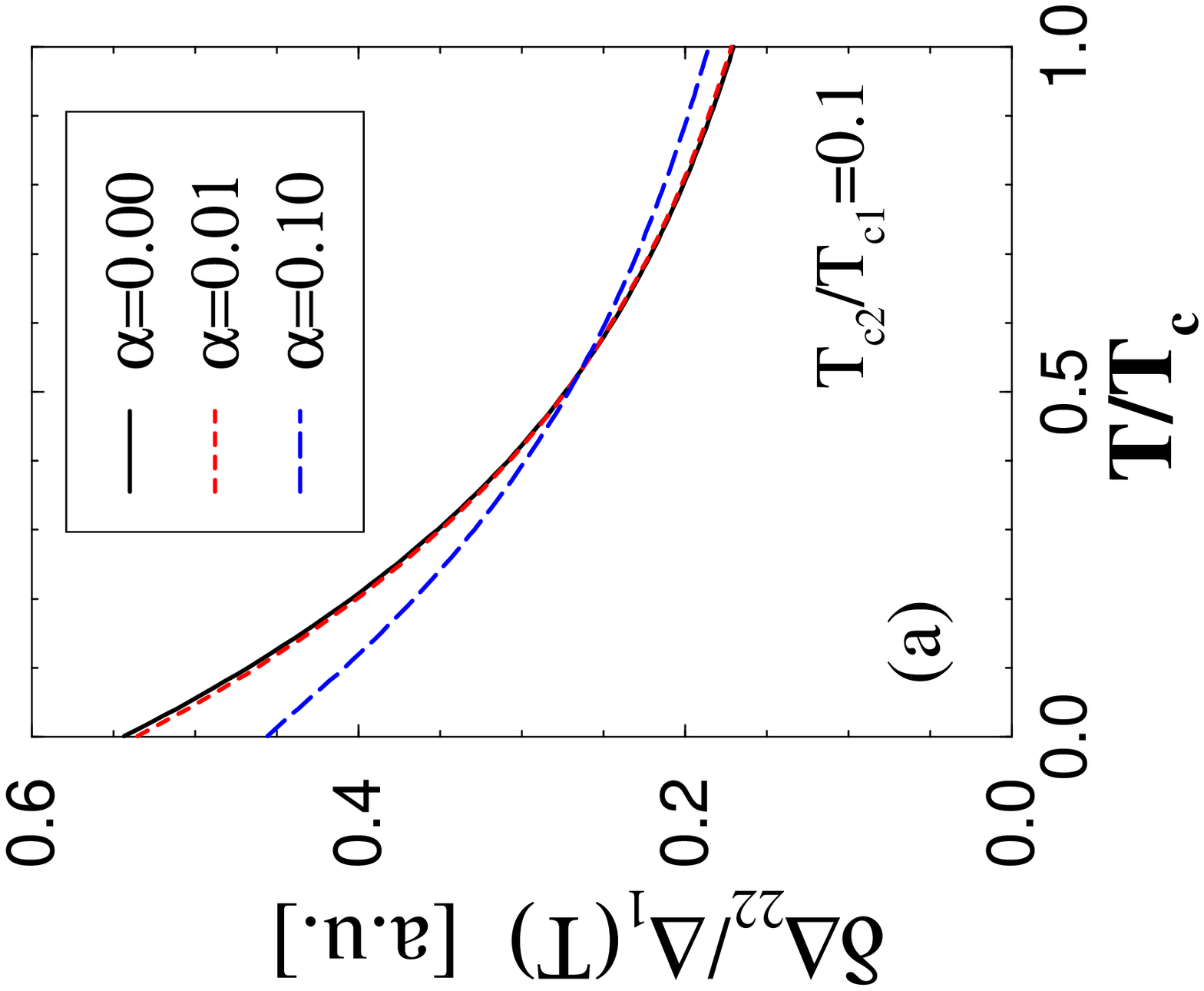}} }
\end{minipage}
\noindent
\begin{minipage}{0.24\textwidth}
\epsfysize=0.99\hsize
\centerline{ \rotate[r]{\epsfbox{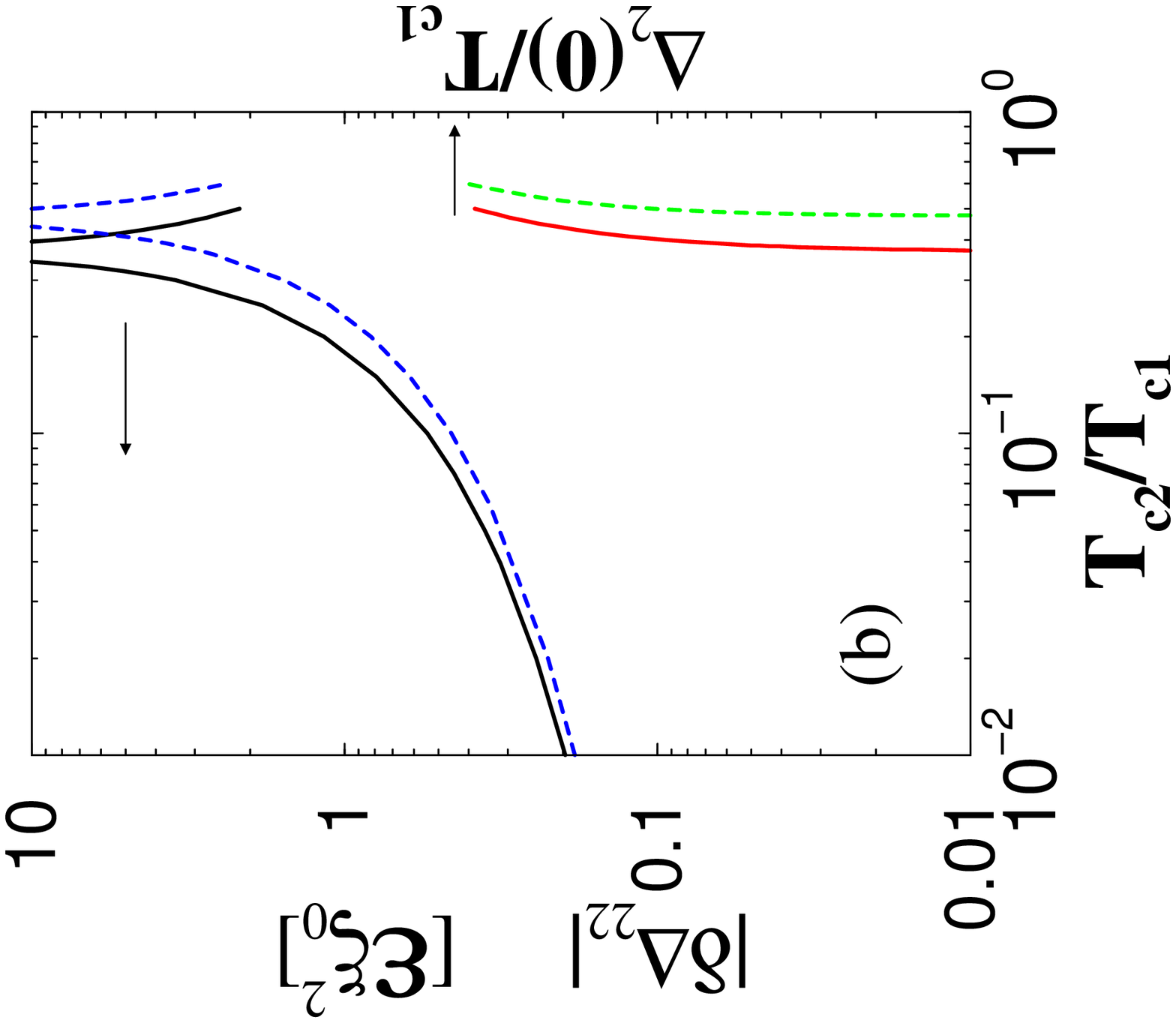}} }
\end{minipage}
\noindent
\begin{minipage}{0.48\textwidth}
\caption[]{Induced $i d_{xy}$ OP at ${\bf q} = 0$.
(a) $\delta\Delta_{22}$ normalized by the $\Delta_1(T)$ OP vs. 
$T_c=T_c(\alpha)$
for different bulk scattering rates $\alpha=1/(2\pi T_{c1} \tau)$. 
(b) $\delta\Delta_{22}$ and the bulk $\Delta_2$ OP
for attractive pairing potentials ($T_{c2}/T_{c1}$) at $T=0$ for
$\alpha = 0.0$ (solid) and $\alpha = 0.1$ (dashed).  
{See text for the definition of ${\cal E}$.}
}\label{2nd}
\end{minipage}
\end{figure}

The corrections to the {bulk} OP,
$\delta\Delta_i$, with $i = 1, 2$, do not belong
to a single representation, {\it i.e.},
$\delta\Delta_i(\phi,{\bf q}) \not\sim \eta_i(\phi)$.
In particular, $\delta\Delta_2(\phi,{\bf q})$ has mixed
$d_{xy}$ and $d_{x^2-y^2}$ symmetry,
$\delta\Delta_2(\phi,{\bf q}) = \delta\Delta_{21}({\bf q})\eta_1(\phi)
+\delta\Delta_{22}({\bf q})\eta_2(\phi)$.
These amplitudes also determine the current
distribution (\ref{current_density}), and satisfy the coupled 
equations
\begin{eqnarray}
\left[ 1/V_1 - {\cal K}_{11}({\bf q}) \right] \delta\Delta_{21}({\bf q}) 
 - {\cal K}_{12} \delta\Delta_{22}({\bf q})
&=& {\cal A}_1({\bf q}) \,,
\\
\left[ 1/V_2 - {\cal K}_{22}({\bf q}) \right] \delta\Delta_{22}({\bf q}) 
 - {\cal K}_{12} \delta\Delta_{21}({\bf q})
&=& {\cal A}_2({\bf q}) \,,
\end{eqnarray}
where
${\cal K}_{ij}({\bf q}) = \pi T\sum_{\epsilon_n} \int \frac{d\phi}{2\pi}
 \eta_i(\phi)\eta_j(\phi) E / [ E^2 + Q^2 ]$, and
${\cal A}_i({\bf q}) = \pi T\sum_{\epsilon_n} t_2(\epsilon_n) 
\int \frac{d\phi}{2\pi} \eta_i(\phi) \eta_2(\phi) E / [ E^2 + Q^2 ]$.
{The solutions to these equations have,
in addition to inversion symmetry,}
{the mirror reflections
$\delta\Delta_{2i}(q_1,q_2)=(-)^i\delta\Delta_{2i}(q_1,-q_2)\!
 = \!(-)^i\delta\Delta_{2i}(q_2,q_1)$.}
The induced {$\delta\Delta_{22}$} OP component with B$_{\text{2g}}$ 
symmetry is finite for ${\bf q}\!=0\!$,
while the induced {$\delta\Delta_{21}$} component with B$_{\text{1g}}$ 
symmetry vanishes for ${\bf q}\!=\!0$
and along the diagonals and principle axes.
The Fourier transformation of $\delta\Delta_{21}({\bf q})$
also vanishes, {\it i.e.}, $\delta\Delta_{21}({\bf r})\!=\!0$; thus only the
induced component with B$_{\text{2g}}$ symmetry survives.
The contour plot of $\delta\Delta_{2}({\bf r})$ 
in Fig.~1 shows a four-fold pattern
characteristic of the $d_{xy}$ amplitude
{with maxima located at approximately}
{$0.3\xi_0$ along the nodal directions}
of the $d_{x^2-y^2}$ OP.

{Bulk impurity scattering is pair-breaking for any unconventional
{OP} including the induced $d_{xy}$ amplitude. 
Figure~\ref{2nd} shows both the temperature dependence of the
induced $d_{xy}$ OP and the pair-breaking suppression
of $\delta\Delta_{22}({\bf q}=0)$ by bulk impurity scattering.
Note that the $d_{xy}$ OP develops below $T_c$ and that it
is suppressed by bulk scattering on the same scale as the
bulk $d_{x^2-y^2}$ OP.
{Fig.~\ref{2nd}(b) shows the increase in the
induced $d_{xy}$ amplitude with increasing (attractive) pairing
interaction in the $B_{2g}$ channel ($T_{c2}/T_{c1}$);
the divergence at $T_{c2}/T_{c1}>0.37$} corresponds to a
bulk instability for $d_{x^2-y^2}\pm id_{xy}$ pairing.}
{For a repulsive pairing interaction $V_2 < 0$ we find
a cutoff dependent result for the induced OP,
$-0.44\,{\cal E}\xi_0^2/\log(\omega_c/\Delta_1) < 
\delta\Delta_{22}(q\!=\!0) < 0$, at $T\!=\!0$.
We parameterized the interference term in the scattering
amplitude by the coupling energy
${\cal E}=\pi S_z N_f \tilde\lambda^2/\xi_0^2$,
}

The existence of a $\pm i d_{xy}$ OP implies that the
equilibrium superconducting state breaks ${\cal T}$
symmetry {\sl locally} near the impurity, in addition
to broken [110] and [100] reflection symmetries.
The signature of the
$d_{x^2-y^2}\pm i d_{xy}$ state is the equilibrium
charge current and magnetic field distribution 
near the impurity.\cite{rainer77,choi89,choi90,salkola98}
From the $\hat t$ matrix in Eq.~(\ref{tmatrix}) and the induced $d_{xy}$ OP
we obtain

\begin{figure}
\noindent
\begin{minipage}{0.48\textwidth}
\epsfysize=0.99\hsize
\centerline{ \rotate[r]{\epsfbox{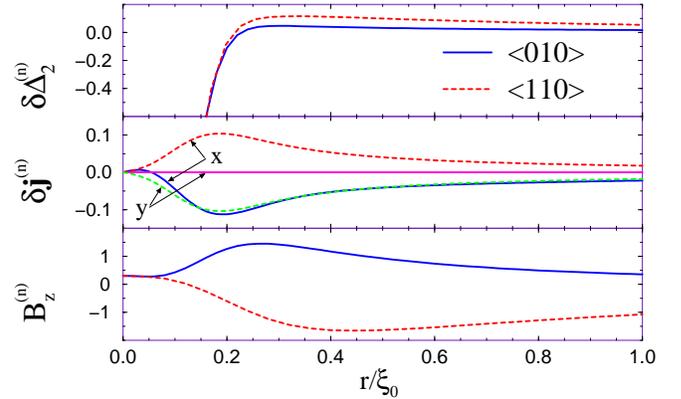}} }
\caption[]{{Spatial dependence of the normalized OP,
$\delta\Delta_2^{(n)}({\bf r})$,
current density, $\delta {\bf j}^{(n)}({\bf r})$ ($x$ and $y$ components), 
and magnetic field, $B_z^{(n)}({\bf r})$, along the 
{$\langle 010 \rangle$ and $\langle 110 \rangle$ directions at $T=0.1 T_c$ 
for a modest $d_{xy}$ pairing interaction of $T_{c2}/T_{c1}=0.1$
(well below the bulk transition to $d_{x^2-y^2}\!+\!id_{xy}$).}}
}\label{tricon}
\vspace{10pt}
\end{minipage}
\end{figure}

\begin{eqnarray}\label{current}
&&
\delta {\bf j}({\bf q}) =
{i \pi e} N_f T \sum_{\epsilon_n} \int \frac{d\phi}{2\pi}
{\bf v}_f(\phi)\, {\bf v}_f(\phi)\!\cdot\!{\bf q}
\frac{ \Delta_1 \eta_1(\phi) }{ E ( E^2 + Q^2 ) }
\nonumber\\ && \qquad \times
\Big(
 \big( t_2(\epsilon_n) + \delta\Delta_{22}({\bf q}) \big) \eta_2(\phi) 
 + \delta\Delta_{21}({\bf q})\eta_1(\phi)
\Big)\,.
\end{eqnarray}

{It is straightforward to show that the current density obeys
the symmetry relations: $\delta{\bf j}({\bf q}) = - \delta{\bf j}(-{\bf q})$, and}
$\delta j_{i}(q_1,q_2) = (-)^{i+1} \delta j_{i}(-q_1,q_2)$,
$\delta j_1(q_1,q_2) = - \delta j_2(q_2,q_1)$.
From these relations one might expect a simple circulation 
pattern for the induced charge current; however, 
as Figs.~1 and \ref{tricon} show, the current density exhibits
the superposition of a {very small} circulating current loop 
and four counter circulating currents around the {nodal directions,}
which are anchored to the local maxima of the 
induced $\delta\Delta_{2}({\bf r})$ OP.
This pattern is qualitatively similar to 
the current distribution predicted for 
a vortex with $d\!+\!is$ symmetry,\cite{franz96}
however, there is no circulation at large 
{distances} from the impurity. The complex flow 
pattern that is observed near the impurity is
also observed for mesoscopic superconductors
with surfaces that are normal to the 
{$\langle 110 \rangle$  direction},\cite{mar98} and reflects
the strong nonlocality of the
current response shown in Eq.~(\ref{current}).

The spatial pattern of current generates a four-fold 
magnetic field distribution which we calculated from the 
current distribution using the  Biot-Savart law,
$B_z({\bf r})={1\over c}\int d^2{\bf r}'
{|{\bf r}'-{\bf r}|^{-3}}{({\bf r}'-{\bf r})\times\delta{\bf j}({\bf r}')}$.
{The field distribution in Fig.~1 shows 8 sectors of flux threading
in and out of the plane.}
{As a result the net magnetic
flux through the superconducting plane is zero. This
fact was checked numerically.}
The magnetic flux was calculated for {squares} of area $L^2$
and shown to vanish in the limit $L \to \infty$.
{This is a general result,}
at least at the quasiclassical level, provided that scattering
by the impurity does not generate a coupling of the current 
to a soft mode of the {OP}.\cite{choi89}
{However, particle-hole asymmetry corrections to the
quasiclassical theory may lead to a net moment from
the impurity-induced orbital currents.}

The magnitudes of the induced $i d_{xy}$ OP 
and magnetic field near an
impurity depend on parameters characterizing the
interaction between quasiparticles and the magnetic
impurity. Not much is known about these interactions. 
Thus, measurements of the induced OP or magnetic field
near an impurity can provide direct information about the
coupling of the quasiparticle orbital momentum to the 
impurity moment. We can express the impurity induced OP,
current density and magnetic field in terms of a few 
key material parameters of the impurity.
We scale the induced OP in units of the coupling energy,
$\delta\Delta_2^{(n)}({\bf r})=\delta\Delta_2({\bf r})/{\cal E}$.
From Eq.~(\ref{current}) we obtain the scale of the 
current and field:
$\delta{\bf j}^{(n)}({\bf r}) = \delta{\bf j}({\bf r}) / (c {\cal B})$,
$B_z^{(n)}({\bf r}) = B_z({\bf r}) / {\cal B}$,
{with ${\cal B}={e\over c} N_f v_f {\cal E} = \frac{\Phi_0}{4\pi\lambda^2}
 \frac{{\cal E} d}{\pi \hbar v_f}$.}
Figure~\ref{tricon} shows the spatial variations of the scaled
OP, current density and field profile along the 
{$\langle 110\rangle$ and $\langle 010\rangle$}
directions at low temperature.

We estimate the magnitude of the induced OP and magnetic
field for Ni impurities in Bi-2212 as follows:
The coupling parameter for Ni$^{2+}$ 
ions is estimated from the spin-orbit coupling energy for free Ni$^{2+}$ 
ions,\cite{yosida96}
{$\tilde\lambda^2 \sim (30 \,{\rm meV} \cdot a_0^2 )^2$. 
The in-plane penetration depth of $\lambda \approx 200\, {\rm nm}$,
the interlayer spacing, $d \approx 1.5\, {\rm nm}$,
the in-plane lattice constant, $a_0 \approx 5.4\, {\rm \AA}$, 
and the Fermi velocity, $v_f \sim 100 {\rm km/s}$,  
provide a determination of the density of
states per Cu-O bilayer, $N_f={c^2 d}/({4\pi}e^2 v_f^2\lambda^2)$. 
This gives an estimated energy scale for the induced
$d_{xy}$ gap of ${\cal E} \sim 0.1 \,{\rm meV}$, and
a magnetic field scale of order ${\cal B} \sim 1 \,{\rm \mu T}$.
An induced $d_{xy}$ gap of order $0.1\, {\rm meV} \approx 1\,{\rm K}$,
is the right order of magnitude to account for a $d_{x^2-y^2} \pm id_{xy}$
phase order transition at lower temperature,
$T^*\sim 150 \,{\rm mK}$, as observed in {0.6\%  and 1.5\%}
Ni doped Bi-2212.\cite{movshovich98,balatsky98}
}

In conclusion, we have shown that spin-orbit
scattering induces a $d_{x^2-y^2}+i d_{xy}$ state,
which locally breaks ${\cal T}$ symmetry
in the vicinity of a magnetic impurity. The induced OP
develops below $T_c$ and survives bulk impurity scattering
so long as the bulk $d_{x^2-y^2}$ OP does.
The signature of the spontaneously broken ${\cal T}$ symmetry 
manifests itself as a complex pattern of circulating charge currents
near the local maxima of the $d_{xy}$ OP located along the
{$\langle 110 \rangle$} and
{$\langle \bar{1}10 \rangle$} directions. 
We estimated the magnitude of the induced
$d_{xy}$ gap to be $\sim 1\, {\rm K}$, which should be observable
in low temperature STM measurements of the tunneling density of states.

We thank R. Movshovich, D. Rainer and  J.R. Schrieffer
for discussions. The work of MJG and AVB was supported by the 
Los Alamos National Laboratory under the auspices of the US
Department of Energy. JAS acknowledges support by the
STCS through NSF DMR-91-20000.

\end{multicols}

\end{document}